\documentclass[aps,prb,twocolumn,showpacs,longbibliography,superscriptaddress,final,floatfix]{revtex4-1}
\usepackage{amsmath}
\usepackage{amsfonts}
\usepackage{amssymb}
\usepackage{physics}
\usepackage{color}
\usepackage{hyperref}
\usepackage[]{inputenc}
\usepackage{enumerate}
\usepackage{graphicx}
\usepackage{float}
\usepackage{amsmath}
\usepackage{amsfonts}
\usepackage{lipsum}
\usepackage{dcolumn}
\usepackage{hyperref}
\usepackage{subfigure}
\usepackage{epsfig}
\usepackage{epsf} 
\usepackage{epstopdf}

\usepackage{color}
\usepackage{enumitem}

\usepackage{xr}
\makeatletter

\begin{document}


\title{ Topologically protected  subdiffusive transport in two-dimensional fermionic wires }


\author{Junaid Majeed Bhat}


\affiliation{Department of Physics, Faculty of Mathematics and Physics, University of Ljubljana, 1000 Ljubljana, Slovenia}
\affiliation{International Centre for Theoretical Sciences, \\
	Tata Institute of Fundamental Research, Bengaluru, 560089, India}


\date{\today}

\begin{abstract}
	The conductance at the band edges of one-dimensional fermionic wires, with $N$ sites, has been shown to have subdiffusive $(1/N^2)$ behavior. We investigate this issue in two-dimensional fermionic wires described by a hopping model on an $N_x\times N_y$ rectangular lattice comprised of vertical chains with a Hermitian intra-chain and inter-chain hopping matrices given by $H_0$ and $H_1$, respectively. We study particle transport using the non-equilibrium Green's function formalism, and show that the asymptotic behavior of the conductance, $T(\omega)$, at the Fermi level $\omega$, is controlled by the spectrum of a dimensionless matrix $A(\omega)=(-\omega+H_0)H_1^{-1}$. This gives three simple conditions on the spectrum of $A(\omega)$ for observing ballistic, subdiffusive, and exponentially decaying $T(\omega)$ with respect to $N_x$. We show that certain eigenvalues of $A(\omega)$ give rise to subdiffusive contributions in the conductance, and correspond to the band edges of the isolated wire. We demonstrate that the condition for observing the subdiffusive behavior can be satisfied if $A(\omega)$ has nontrivial topology. In that case, a transition from ballistic behavior to subdiffusive behavior of the conductance is observed as the hopping parameters are tuned within the topological regime. We argue that at the transition point, different behaviors of the conductance can arise as the trivial bulk bands of $A(\omega)$ also contribute subdiffusively. We illustrate our findings in a simple model by numerically computing the variation of the conductance with $N_x$. Our numerical results indicate a different subdiffusive behavior ($1/N_x^3$) of the conductance at the transition point. We find the numerical results in good agreement with the theoretical predictions.	
\end{abstract}

\maketitle
\section{Introduction}
 Electronic transport in mesoscopic systems has been extensively explored for several decades now~\cite{datta1997electronic,imry1997introduction,Buttiker1985Gen,dhar2006}. However, new and interesting results are discovered in very simple models even after decades of research 
 . One example would be the exotic transport characteristics of topological insulators and superconductors~\cite{PhysRevLett.118.236402,liu2022chern,peng2021topological,roy2012,bondyopadhaya2020nonequilibrium,Bhat2020transport} which have taken a central stage in modern quantum condensed matter~\cite{shankar2018,hasan2010colloquium,moore2010birth,PhysRevLett.95.146802,PhysRevLett.98.106803}. They are of interest as they possess the ability to carry dissipationless current via robust edge modes. These modes are protected from  symmetry-preserving weak disorders as they are gapped from the bulk bands of the system. In one dimension,  the simplest examples of a topological insulator and  a topological superconductor are the Su-Schrieffer-Heeger model (SSH)~\cite{su1979solitons,PhysRevB.100.075437},  which describes poly-acetylene, and the Kitaev chain~\cite{kitaev}, respectively.  In two dimensions  the Haldane model~\cite{PhysRevLett.61.2015}, Bernevig, Hughes and Zhang model~(BHZ)~\cite{bernevig2006},  and Kane-Mele model~\cite{PhysRevLett.95.146802} are known to posses non trivial band topology. Some of these models have been realized in different experimental setups~\cite{jotzu2014experimental,konig2007}. 

 Another very recent example is the work of  Refs.~\onlinecite{PhysRevLett.130.187101,PhysRevB.108.075406} on the scaling of the conductance with system size in 1D fermionic wires at the band edges of the wire. It was found that a universal subdiffusive behavior of the conductance is observed if the Fermi level of the reservoirs is set at the band edges of  the isolated fermionic wire. Therefore, the subdiffusive behavior separates the transition from ballistic to exponential decay of the conductance as one tunes the Fermi level of the reservoirs from inside to outside the energy band. The band-structure of nearest neighbor  fermionic wire with translation symmetry  can  obtained from $2\times2$ transfer matrices~\cite{last1993relation,molinari1997transfer,molinari1998transfer} and the subdiffusive behavior arises from the non-diagonalizibilty of this matrix. Similar peculiarities in the behavior of the  wave functions near the band edges have been reported in earlier works also~\cite{dunlap1990absence,pryor1992electron,kawarabayashi1995diffusion}.  
 
 In this paper, we show that the interplay of the above two examples reveals very interesting physics in two dimensions.  We first look at the problem of the scaling of the conductance at the band edges in two dimensions, which to our knowledge has not been explored so far. To do that, we consider a fermionic wire described by a hopping model on a rectangular lattice ($N_x\times N_y$). We think of the wire to be comprised of $N_x$ vertical chains of length $N_y$, and  for simplicity we assume  a Hermitian intra-chain as well as  inter-chain hopping matrices $H_0$ and $H_1$, respectively. We use the non-equilibrium Green's function formalism (NEGF)~\cite{datta1997electronic,dhar2006,roy2012,Bhat2020transport} formalism to explore the asymptotic behavior of conductance at the Fermi level $\omega$ with respect to $N_x$, and  find that this behavior  is controlled by the spectrum of a matrix, $A(\omega)=(-\omega+H_0)H_1^{-1}$. The Hermiticity of $H_0$ and $H_1$,  implies $A(\omega)$ is pseudo Hermitian with respect to $H_1$.  Such matrices have been found important in different problems of physics~\cite{wigner1960normal,pauli1943dirac,cardy1985conformal,dorey2001spectral,mostafazadeh2002pseudo,mostafazad2002pseudo,fisher1978yang,ashida2020non}. Our work uncovers an interesting  connection of such matrices with the transport in two-dimensional fermionic wires.

 We find that the real eigenvalues of $A(\omega)$ of magnitude $2$ lead to subdiffusive ($1/N_x^2$) contributions in the conductance, and occur with the Fermi level at the band edges of the isolated wire.  Other real eigenvalues of magnitudes less than  $2$ give rise to ballistic contribution in the conductance, and correspond to Fermi level within the energy bands of the isolated wire. The remaining eigenvalues of $A(\omega)$ contribute exponentially small in the conductance.
 
 The condition for observing  subdiffusive behavior  reads that $A(\omega)$ has at least  one eigenvalue of magnitude $2$, and simultaneously all the other real eigenvalues  greater than $2$. This condition implies that the subdiffusive behavior would be seen only at the band edges of the isolated wire that do not lie within any other bands of the system. Therefore, as the positions of other bands of the isolated wire are tuned with respect to the edge of a particular band a ballistic to subdiffusive transition in the behavior of the conductance occurs.  Since the  Fermi level is kept fixed at some band edge, this band edge should remain fixed as the  positions of the other bands are tuned by varying the parameters of $H_0$ and $H_1$. We present a toy example in $2\times N_x$ ladders  where this transition occurs.
 
  Interestingly, we find that if $A(\omega)$ is topologically nontrivial, then the subdiffusive condition can be satisfied and a transition  from ballistic to subdiffusive ($1/N_x^2$) behavior occurs within the topologically nontrivial parameter regimes. In this case, the subdiffusive behavior is due to the topological modes, and therefore it possesses the same topological protection from effects of disorder as the  modes themselves. In contrast, for the parameter regimes where  $A(\omega)$ is topologically trivial, a usual transition from ballistic to exponential decay of the conductance exists. We refer to the two transitions as the subdiffusive and exponential transitions, respectively.  At the transition points,  we argue novel behaviors of the conductance can arise for thermodynamically large $N_y$ as at these points the bulk bands of $A(\omega)$ also contribute subdiffusively.

 We illustrate our theoretical expectations by considering a simple model comprised of SSH chains. We numerically compute the behaviors of the conductance  for this model using an iteration method  based on the Dyson equation for the Green's functions~\cite{mahanmany,metalidis2007electronic}.   Our numerical computations of the conductance  agree with the theoretical expectations away from the transition points. At the transition points, our numerical computations indicate a different subdiffusive behavior $(1/N_x^3)$ in the limit of $N_y\rightarrow\infty$. At finite $N_y$, we see a crossover from $(1/N_x^3)$ to $(1/N_x^2)$, and to exponentially decaying conductance for the subdiffusive and exponential transition point, respectively.

This paper is structured as follows: Sec.~\ref{sec:model} introduces the details of the wire Hamiltonian,  the Hamiltonian of the reservoirs, and their couplings with the wire. In the next section, Sec.~\ref{sec:cond}, we discuss a simplification of the NEGF conductance formula from which its asymptotic behavior with $N_x$ become explicit in terms of the spectrum of $A(\omega)$. In the next section Sec.~\ref{sec:condnx}, we  discuss the conditions for observing the different behaviors of the conductance namely ballistic, subdiffusive, and exponential decay. In  section Sec.~\ref{sec:tpl}, we discuss how the subdiffusive transition can occur in the parameters of the topological regime, if $A(\omega)$ has a nontrivial topology.  In the penultimate section, we look at a simple model and discuss the subdiffusive and the exponential transitions in this case. We also discuss the effects of disorder on the subdiffusive behavior. We  then  present numerical results for the conductance of this model at $\omega=0$ to confirm our expectations from the theory. We also present numerical results on the conductance at the transition points, where we find a $1/N_x^3$ subdiffusive behavior for thermodynamically large $N_y$. We conclude in Sec~\ref{sec:concl}.

\section{Model}
\label{sec:model}
We consider the wire Hamiltonian, $\mathcal{H}_W$, to be given by a tight-binding model defined on a rectangular lattice of size $N_x\times N_y$.  The left edge  and the right edge of the wire is placed in contact with   reservoirs. These are taken to be  semi-infinite metallic leads  of width $N_y$ and are modeled as 2D tight-binding Hamiltonians which we refer to as $\mathcal{H}_L$ and $\mathcal{H}_R$, respectively.  We label the  annihilation and creation operators at a site $(x,y)$ as ($\psi(x,y)$, $\psi^\dagger(x,y)$), and $(\phi_{L/R}(x,y), \phi^\dagger_{L/R}(x,y))$ on the wire, the left lead and the right lead, respectively. These satisfy the usual Fermionic anti-commutation relations. Free boundary conditions are imposed at the horizontal edges of the reservoir and the system at $y=1$ and $y=N_y$, respectively. The contacts between the wire and the reservoirs are themselves modeled as tight-binding Hamiltonians, $\mathcal{H}_{WL}$ and $\mathcal{H}_{WR}$. The full Hamiltonian of system  given by,
\begin{equation}
\mathcal{H}=\mathcal{H}_L+ \mathcal{H}_{LW}+ \mathcal{H}_W+ \mathcal{H}_{RW}+\mathcal{H}_R,
\end{equation}
where the individual Hamiltonians of the wire, the contacts, and the reservoirs  are given by,
\begin{align}
\mathcal{H}_W&=\sum_{x=1}^{N_x} \Psi^\dagger(x)H_0\Psi(x)+  \sum_{x=1}^{N_x-1}\left[\Psi^\dagger(x)H_1\Psi(x+1)+ \text{h.c.}\right],\\
\mathcal{H}_{LW}&= \eta_c (\Psi^\dagger(1)\Phi_L(0)+\Phi_L^\dagger(0)\Psi(1)),\\
\mathcal{H}_{RW}&= \eta_c (\Psi^\dagger(N_x)\Phi_R(N_x+1)+\Phi_R^\dagger(N_x+1)\Psi(N_x)),\\
\mathcal{H}_{L}&=\sum_{x=-\infty}^{0} \Phi^\dagger_L(x)\tilde H_{0}\Phi_L(x)\notag\\&+\eta_{bx} \sum_{x=-\infty}^{-1}\Phi_L^\dagger(x)\Phi_L(x+1)+ \Phi_L^\dagger(x+1)\Phi_L(x),
\end{align}
\begin{align}
\mathcal{H}_{R}&=\sum^{\infty}_{x=N_x+1} \Phi^\dagger_R(x)\tilde H_{0}\Phi_R(x)\notag\\&+\eta_{bx} \sum^{\infty}_{x=N_x+1}\Phi_R^\dagger(x)\Phi_R(x+1)+ \Phi_R^\dagger(x+1)\Phi_R(x).
\end{align}
\begin{figure}
	\subfigure{
		\includegraphics[width=0.49\textwidth]{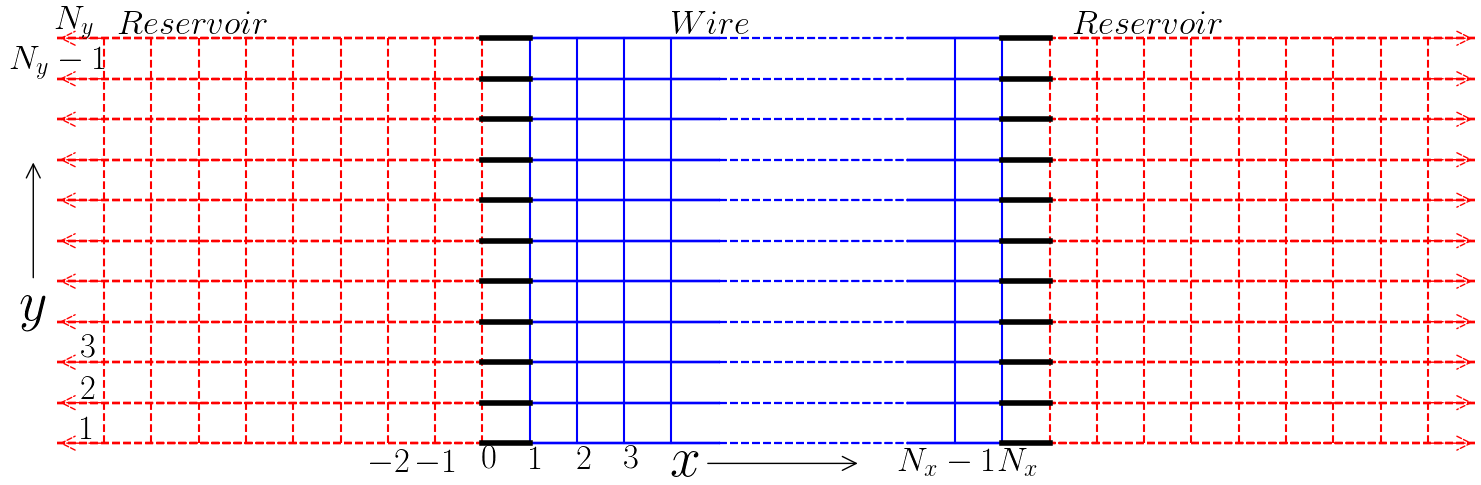}}
	\caption{Schematic of the wire in contact with the reservoirs along the left and right edges.   This figure corresponds to the case where the hopping matrices, $H_0$, $H_1$, and $\tilde H_0$ allow only nearest neighbor hopping between the sites of the entire system.}
	\label{scfig}
\end{figure}
We have defined column vectors $\Psi(x)$, and $\Phi_{L/R}(x)$ of $N_y$ components  with the $y^{th}$ component given by the operators $\psi(x,y)$ and $\phi_{L/R}(x,y)$, respectively. These define the annihilation operators of a single chain of length $N_y$ located at the position $x$. $H_0$, $\tilde H_0$ are the intra-chain hopping matrices inside the wire and the reservoirs, repectively. The inter-chain hopping matrix for the wire is taken to be $H_1$, while the same is assumed to be $\eta_{bx}I$ ($I=$ identity matrix) for the reservoirs. Note that we have taken the reservoir Hamiltonians to be the same, and $H_1$ is assumed to be Hermitian and invertible. The couplings at the contacts are assumed to be of strength $\eta_c$. A schematic of the geometry with $H_0$, $H_{1}$, and $\tilde H_{0}$ which only allow hoppings between nearest neighbor sites of the  system is shown in Fig.~\ref{scfig}. 

Before moving to the next section to discuss the conductance of the wire within the NEGF formalism, it is instructive to note that the bands of the isolated wire for $N_x\rightarrow \infty$ and finite $N_y$ are given by diagonalizing the matrix $H_0+2H_1\cos(k_x)$ where $k_x\in (-\pi,\pi)$ is the wave vector in the $x$ direction. Therefore, the wire can have a maximum of $N_y$ number of bands.

\section{Conductance in the thermodynamic limit}
\label{sec:cond}
Using the NEGF formalism we can determine the nonequilibrium steady state (NESS) of the wire from which all the steady properties including the conductance is computed. The NESS is obtained starting from an initial state where the left and the right reservoirs described by grand canonical ensembles at  temperatures $T_L$ and $T_R$ and chemical potentials $\mu_L$ and $\mu_R$, respectively, and the wire is chosen to be in an arbitrary state. It is then described in terms of the effective nonequilibrium Green's function for the wire defined as~\cite{dhar2006, Bhat2020transport},
\begin{equation}
G^+(\omega)=\frac{1}{\omega-H_W-\Sigma_L(\omega)-\Sigma_R(\omega)},
\end{equation}
where $\Sigma_L(\omega)$ and $\Sigma_R(\omega)$ are the self energy contributions due to the reservoirs, and $H_W$ is the full hopping matrix of the wire given by,

\begin{equation}
H_W=\begin{pmatrix}
H_0  & H_1&0  &0& \dots\dots& 0&0 \\
H_1& H_0 &  H_1&0 & \dots\dots&0&0 \\
0 & H_1 & H_0 & H_1 &\dots \dots &0&0\\
\vdots&\vdots &\vdots & \ddots&&\vdots&\vdots \\
\vdots&\vdots&\vdots& & \ddots &\vdots&\vdots \\
\vdots&\vdots&\vdots&\dots& H_1&H_0 & H_1\\
0&0 & 0&\dots&\dots&H_1 & H_0 
\end{pmatrix}.
\end{equation}
Due to the contacts only at the edges of the wire, the only nonzero components of $\Sigma_{L}$ and $\Sigma_{R}$ are given by $\Sigma[y,y']=\Sigma_{L}[1,y;1,y']=\Sigma_{L}[N_x,y;N_x,y']$. These  components are the same because the left and right are chosen to have the same Hamiltonian. It can be shown that $\Sigma=U_L \Sigma_D  U_L^\dagger$, where $\Sigma_D$ is a diagonal matrix  with  components given by~\cite{dhar2006},

	\begin{align}
	\Sigma_D[k,k]=\begin{cases}
	\frac{\eta_c^2}{\eta_{bx}}[z_k-i \sqrt{1-z_k^2}]; & |z_k|<1,\\
	\frac{\eta_c^2}{\eta_{bx}}[z_k- \sqrt{z_k^2-1}]; & z_k>1,\\
	\frac{\eta_c^2}{\eta_{bx}}[z_k+ \sqrt{z_k^2-1}]; & z_k<-1,
	\end{cases}
	\label{eq:sed}
	\end{align}
where $z_k=\frac{\omega-\chi_k}{2\eta_{bx}}$. $U_L$ is the unitary transformation that diagonalizes the intra-chain hopping matrix  $\tilde{H}_0$ of the reservoirs and its eigenvalues are given by $\chi_k$, $k=1,2,...,N_y$.
 
With the reservoirs kept at zero temperature and their chemical potentials chosen to have a small bias around the Fermi level  $\omega$, the conductance of the wire, in units of $e^2/h=1$, is given in terms of $G^+(\omega)$ as,
\begin{equation}
T(\omega)=4\pi^2\Tr[G^+(\omega)\Gamma_R(\omega)G^-(\omega)\Gamma_L(\omega)],
\end{equation}
where $G^-(\omega)=[G^+(\omega)]^\dagger$ and $\Gamma_{L/R}=(\Sigma_{L/R}^\dagger-\Sigma_{L/R})/(2\pi i)$. Since the contacts with the reservoirs are only along the edges i.e. at $x=1$, and $x=N_x$, the above trace can be computed as,
\begin{equation}
T(\omega)=4\pi^2 \Tr[G^+_{1N_x}\Gamma G^-_{N_x1}\Gamma],
\label{eq:cond_2}
\end{equation}
where $G^+_{1N_x}[y,y']=G^+(\omega)[x=1,y;x'=N_x,y']$ is the $(1,N_x)$ block of the full Green's function matrix, $G^+(\omega)$ and $G^-_{N_x1}=[G^+_{1N_x}]^\dagger$. Similarly, $\Gamma[y,y']=\Gamma_R[N_x,y;N_x,y']=\Gamma_L[1,y;1,y']$ and these are the only non-zero blocks $\Gamma_L$ and $\Gamma_R$.  

 We need to find the $(1,N_x)$ block, $G^+_{1N_x}$, of the Green's function, $G^+(\omega)$ to determine the conductance. This can be done by using the transfer matrix approach. To that end, we start from the identity $G^+(\omega)(\omega-H_W-\Sigma_L-\Sigma_R)=I$, and exploit the block tridiagonal structure of $H_W$ to obtain the following equations for the inverse of the block components, $G^+_{1i}$, of the Green's function $G^+(\omega)$,
\begin{align}
G^+_{11}(\omega-H_0-\Sigma)+ G^+_{12}H_1&=I,\\
G^+_{1,i-1}H_1+G^+_{1,i}(\omega-H_0)+G^+_{1,i+1}H_1&=0;~1<i<N_x,\\
G^+_{1,N_x-1}(\omega-H_0-\Sigma)+ G^+_{1N_x}H_1&=0,
\end{align}
 Using the transfer matrix approach we can show~\cite{bhat2022heat,cane2021localization,bhat2021equivalence}
\begin{equation}
\begin{pmatrix}I && G^+_{11}\end{pmatrix}=\begin{pmatrix}G^+_{1N_x} && 0\end{pmatrix}\Omega_R\Omega^{N_x}\Omega_L, 
\end{equation}
where,
\begin{align}
\notag\Omega_R=\begin{pmatrix} I && -\Sigma H_1^{-1} \\
0 && 0
\end{pmatrix},&~\Omega_L=\begin{pmatrix} -H_1 && 0\\
-\Sigma && I
\end{pmatrix},\\
\text{and},~&\Omega=\begin{pmatrix} (-\omega+H_0)H_1^{-1} && I \\
-I && 0
\end{pmatrix}.
\end{align} 
Let $U$ be such that $A(\omega)=(-\omega+H_0)H_1^{-1}=U\Lambda(\omega) U^\dagger$, where $\Lambda(\omega)$ is a diagonal matrix comprising of the eigenvalues of $A(\omega)$. Using this we have, 
\begin{align}
\Omega^{N_x}&=\begin{pmatrix}U && 0\\
0 && U 
\end{pmatrix}\begin{pmatrix}
 \Lambda(\omega) && I\\
-I && 0
\end{pmatrix}^{N_x}\begin{pmatrix}U^\dagger && 0\\
0 && U^\dagger 
\end{pmatrix}\\
&=\begin{pmatrix}U && 0\\
0 && U 
\end{pmatrix}\begin{pmatrix}
P_{N_x} && P_{N_x-1}\\
-P_{N_x-1} && -P_{N_x-2}
\end{pmatrix}\begin{pmatrix}U^\dagger && 0\\
0 && U^\dagger 
\end{pmatrix},
\end{align}
where $P_{N_x}$ is a diagonal matrix comprising of diagonal entries given by $p^k_{N_x}$. These obtained via the iteration equation
\begin{equation}
p_{n+1}^k=\lambda_k(\omega)p_{n}^k-p_{n-1}^k;~~~p_0^k=1,~\text{and}~ p_1^k=\lambda_k(\omega),
\label{eq:pn}
\end{equation}
where $\lambda_k(\omega)$ is the $k^{th}$ eigenvalue  of the matrix $A(\omega)$. With this simplification of $\Omega$ we get,
\begin{align}
G^+_{1N_x}=-UG'_{1N_x}U^\dagger,
\label{eq:G1}
\end{align}
 $G'_{1N_x}=[P_{N_x}H_1'+\Sigma'H_1^{'-1}P_{N_x-1}H_1'+P_{N_x-1}\Sigma' +\Sigma'H_1^{'-1}P_{N_x-2}\Sigma' ]^{-1}$. We have expressed $\Sigma$ and $H_1$ in the diagonal basis of $A(\omega)$ as $\Sigma'=U^\dagger\Sigma U$, and $H_1'=U^\dagger H_1U$. This equation can be substituted in Eq.~(\ref{eq:cond_2}) to obtain the conductance of the wire. 

We now assume a weak coupling limit given by $\frac{\eta_c}{\eta_{bx}}<<1$, then the  leading behavior of the $G'_{1N_x}\approx -H_1^{'-1}P^{-1}_{N_x}$ because the self energy terms  are proportional to $\eta_c^2/\eta_{bx}$. Since $P_{N_x}^{-1}$ is diagonal   the conductance simplifies to the following simple expression,
\begin{equation}
T(\omega)\approx4\pi^2\sum_{k,k'} \frac{\mathcal{L}_{k,k'}}{p_{N_x}^k p_{N_x}^{*k'}},\label{eq:condkk}
\end{equation}
where $\mathcal{L}_{k,k'}=\Gamma'_{k,k'}(H_1^{'-1}\Gamma' H_1^{'-1})_{k',k}$. For large $N_y$ the sum over $k,k'$ can be replaced by an integral with an appropriate density, $g(\lambda,\omega)$ of real eigenvalues of $A(\omega)$.  We show in appendix~\ref{app:gencodn} that the complex eigenvalues always contribute an exponentially small amount in the conductance for large $N_x$, so we ignore them here. Therefore, in the continuum limit, we can  write the conductance expression in terms of $g(\lambda,\omega)$ as,
 
\begin{equation}
\label{eq:thlT}
	T(\omega)=4 \pi^2 \int d\lambda d\lambda' g(\lambda,\omega)g(\lambda',\omega)\frac{|\mathcal{L}(\lambda,\lambda')|^2}{p_{N_x}(\lambda)p_{N_x}(\lambda')},
\end{equation}
where we replaced the  dependencies on the indices $k$ and $k'$, with functions of $\lambda$ and $\lambda'$, respectively. 

A natural question to address here is how the density $g(\lambda,\omega)$ is related to the density of states, $g_W(E)$, of the isolated wire in the thermodynamic limit as we would expect the latter to play a role in the conductance expression. We show that these two densities are related to each other for the case where $H_0$ and $H_1$  have translational symmetry and commute. In that case, the spectrum of the isolated wire is given by $\epsilon(k_x, k_y)=\mu_0(k_y)+2 \cos(k_x) \mu_1(k_y)$. $k_x$, $k_y$ being the wave vectors in the $x$ and $y$ direction, respectively, and $\mu_0(k)$ and $\mu_1(k)$ are the spectrum of the matrices $H_0$ and $H_1$, respectively. The density of states  of the wire $g_W(E)$ can be obtained from the following integral,
\begin{align}
	g_W(E)&= \int dk_x dk_y \tilde g_W(k_x,k_y)\\&=\int dk_x dk_y\delta(E-\mu_0(k_y)-2 \cos(k_x) \mu_1(k_y)).\notag
\end{align}	
On the other hand, we can write the eigenvalues of $A(\omega)$ as $\lambda_{k_y}=(-\omega+\mu_0(k_y))/\mu_1(k_y)$, from which the density can be computed by doing the following integral,
\begin{align}	
	g( \lambda,\omega)&=\int dk_y \delta(\omega-\mu_0(k_y)+\mu_1(k_y)\lambda).
\end{align}
On comparing the two equations for the densities, we see  that if $k_x=\arccos(-\lambda/2)$ and $\omega=E$, then
\begin{equation}
g(\lambda, \omega)= \sin(k_x(\lambda))\int dk_y \tilde{g}_W(k_x(\lambda),k_y).\label{eq:denrel}
\end{equation}
So, $g(\lambda,\omega)$ is  the total density of the states of the isolated wire in the thermodynamic limit at a particular $k_x$. The two densities should also  be related  for arbitrary $H_0$ and $H_1$ but it seems nontrivial to relate the two.
\section{Behaviors of the Conductance at large $N_x$}
\label{sec:condnx}
The Hermiticity of $H_0$ and $H_1$ implies that $A^\dagger(\omega)=H_1^{-1}A(\omega)H_1$, and therefore $A(\omega)$ is pseudo Hermitian with respect to $H_1$. It can have complex eigenvalues in general but for this section and the rest of the paper, we assume a real spectrum of $A(\omega)$ to discuss the behaviors of the conductance at large $N_x$. We relegate the case of complex eigenvalues, which isn't very different, to the appendix~\ref{app:gencodn}. 

From Eq.~\ref{eq:condkk}, we note that the behavior of the conductance with respect to $N_x$ is controlled by the $N_y$ number of iteration equations for $p_n^k$ defined at each eigenvalue $\lambda_k(\omega)$ of the matrix $A(\omega)$. The solution for the iteration can be expressed in terms of the $2\times 2$ transfer matrices given by,
\begin{equation}
\begin{pmatrix}
p_{n+1}^k\\ p_n^k
\end{pmatrix}=M^n\begin{pmatrix}
p_1^k\\p_{0}^k
\end{pmatrix},
\end{equation}
where, $M=\begin{pmatrix}
\lambda_k(\omega) && -1\\
1 && 0
\end{pmatrix}$. It is straightforward to see if $\lambda_k(\omega)\neq2$, then $p^k_{N_x}=\sin(q^k (N_x+1))/\sin(q^k)$, where, $q^k=\arccos[\frac{\lambda_k(\omega)}{2}]$. Note that for $|\lambda_k(\omega)|<2$,  $p_n^k$ is oscillatory in $n$ while for $|\lambda_k(\omega)|>2$, $p_n^k$ grows exponentially. If $\lambda_k(\omega)=2$, then the matrix $M$ becomes nondiagonalizable and $p_{N_x}^k=N_x+1$. 
Depending on the behaviors of all the $p_{N_x}^k$'s with $N_x$, which are in turn determined by the spectrum of $A(\omega)$,  we can have three possible behaviors of the conductance in the limit, $N_x\rightarrow\infty$ at finite $N_y$. These are given by,
	\begin{itemize}
		\item Ballistic ($T(\omega)\sim e^{i\alpha N_x}$, $\alpha$  is real): The ballistic contribution comes from $\lambda_k(\omega)'s$ with absolute values less than $2$, since for such eigenvalues $p_{N_x}^k\sim e^{\pm iq^k N_x}$. The conductance is proportional to the inverse of $p_{N_x}^k$, and therefore the ballistic contribution dominates the transport as other contributions  are suppressed in the limit $N_x\rightarrow\infty$. Hence, a ballistic behavior of the conductance  occurs if at least one of the $\lambda_k(\omega)$'s has a  magnitude of less than $2$. This condition is always satisfied  by $\omega$ lying within a band of the bulk spectrum of the wire for arbitrary $H_0$ and $H_1$.  This is readily seen if $[H_0,H_1]=0$,  then $\lambda_k(\omega)=(-\omega+\mu_0^k)/\mu_1^k$, where $\mu_0^k$ and $\mu_1^k$ are the $k^{th}$ eigenvalues of $H_0$ and $H_1$, respectively. Hence, for a particular $k$, $|\lambda_k(\omega)|<2$ gives $E_{min}^k<\omega<E_{max}^k$, where $E_{max}^k$ and $E_{min}^k$ are the maximum and minimum of $\mu_0^k\pm 2 \mu_1^k$, respectively. Note that $E_{max}^k$ and $E_{min}^k$ are exactly the band edges of the $k^{th}$ band of the bulk spectrum, given by $\mu_0^k+2\cos(k_x)\mu_1^k$, of the isolated wire. We consider the case where $H_0$ and $H_1$ do not commute in the appendix~\ref{app:sdproof}. If all the eigenvalues of $A(\omega)$ contribute balistically, we see that the integration over $\lambda$ and $\lambda'$ in Eq.~(\ref{eq:thlT})  can be restricted over the values of with $|\lambda|<2$, and we have,
		\begin{equation}
			T_{b}(\omega)= \int_{-2}^2\int_{-2}^{2} d\lambda d\lambda' g(\lambda,\omega)g(\lambda',\omega) \frac{|\mathcal{L}(\lambda,\lambda')|^2}{p_{N_x}(\lambda)p_{N_x}(\lambda')}.
		\end{equation}
		\item Exponential decay ($T(\omega)\sim e^{-\alpha N_x},~\alpha>0$): This behavior occurs when all the $\lambda_k(\omega)$'s have magnitude greater than $2$. In that case, every $p_{N_x}^k$ will  grow exponentially with $N_x$, and therefore contribute an exponentially small amount in the conductance. A similar analysis as in the ballistic case gives that the exponential decay occurs only at $\omega>E_{max}^k$ or $\omega<E_{min}^k$   for  $[H_0,H_1]=0$. Therefore, we see exponentially decaying behavior  outside the energy bands of the isolated wire. For this case the leading behavior of the conductance is given by the eigenvalue, $\lambda_{min}$ with minimum $\Im[\arccos(\lambda_{k}/2)]$. Using this fact in Eq.(\ref{eq:thlT}), we have 
		\begin{equation}
		T_e(\omega)\sim g^2(\lambda_{min},\omega)e^{-\alpha N_x}, \label{eq:exp_cond}
		\end{equation}
		where $\alpha=2 |\Im[\arccos(\lambda_{min}/2)]|$.
	
	
\item Subdiffusive ($T(\omega)\sim1/N^2_x$):	The subdiffusive contribution comes from $\lambda_k(\omega)'s$ of absolute value equal to $2$ as for such eigenvalues $p_{N_x}^k\sim N_x$, and therefore their contribution in conductance is proportional to $1/N_x^2$. Hence, the condition for observing the subdiffusive behavior is that at least one of the $\lambda_k(\omega)$'s should have a magnitude of $2$ and at the same time, all the other eigenvalues should be of magnitude greater than $2$. The latter part of the condition is to ensure that there are no eigenvalues which contribute ballistically. Otherwise the contribution from such eigenvalues will dominate the behavior of the conductance at large $N_x$. 
Once again if $[H_0, H_1]=0$, then  $\lambda_{k^*}(\omega)=2$ implies that $\omega$ is  the same as the band edges, $E_{max}^{k^*}$ or $E_{min}^{k^*}$, of the isolated wire. In the appendix~\ref{app:sdproof}, we show that this is true even when $H_0$ and $H_1$ do not commute. Note that  the subdiffusive condition  will only be satisfied at the band edges, that do not lie within any other energy bands of the spectrum of the isolated wire. The subdiffusive contribution of the band edges lying within any other band would be subdued due to the ballistic contribution of the other band. For the subdiffusive case, only the modes at $\lambda=2$ contribute in Eq.~\ref{eq:thlT} and therefore,
	 \begin{equation}
	 	T_{s}(\omega)\sim \frac{g^2(2,\omega)}{p_{N_x}^2(2)}\sim g^2(2,\omega) \frac{1}{N_x^2} .
	 \end{equation}
\end{itemize}	
 We refer to the eigenvalues of $A(\omega)$ that give rise to the three different behaviors as ballistic, exponential, and subdiffusive eigenvalues, respectively.  The fact that the subdiffusive behavior is only seen when the band edge does not lie inside any other energy band of the wire, suggests  that a transition in the behavior of the conductance from ballistic to subdiffusive behavior can arise as the position of the other bands is changed with respect to the band edge where the Fermi level is set. However, for the Fermi level fixed at some band edge, this edge should remain uneffected as the  positions of the other bands are tuned by varying the parameters of $H_0$ and $H_1$. We present a toy example in $2\times N_x$ ladders in the appendix~\ref{app:toy} where this occurs. In the next section, we discuss  that this transition also occurs if  $A(\omega)$ has  nontrivial topology which is more interesting as we would then expect the subdiffusive behavior to be protected from the effects of weak disorder in $A(\omega)$. 
\section{Topologically non-trivial $A(\omega)$}	 
\label{sec:tpl}
  Let us order $\lambda_{k}(\omega)$ such that $\lambda_1<\lambda_2<\lambda_3<...<\lambda_{N_y}$. As per the subdiffusive condition, we must have for some $k=k^*$, $|\lambda_{k^*}|=2$ and simultaneously $|\lambda_{k\neq k^*}|>2$.  If $\lambda_{k^*}=2$, then latter part of the condition is not a problem for $k>k^*$, those eigenvalues are by definition bigger than $\lambda_{k^*}$. In order to have $|\lambda_{k<k^*}|>2$, $\lambda_{k^*}$ should be gapped from below with a gap strength of at least $4$. Similar arguments for $\lambda_{k^*}=-2$ would give that it should be gapped from above. We fix $\lambda_{k^*}=2$ to avoid any confusions. This required structure in the spectrum of the matrix $A(\omega)$ has a striking similarity with 1D insulators of nontrivial topology where the gap could be tuned keeping the topological modes at a fixed eigenvalue.
  
   Let us now fix $\omega=0$ for simplicity, so we choose $A(\omega=0)$ to have a non-trivial topology and fix $\lambda_{k^*}(\omega=0)=2$ to correspond to a topological mode.  Then a transition from ballistic to subdiffusive behavior in the conductance at $\omega=0$ will take place as the gap between the topological mode and the  other eigenvalues  becomes greater than $4$.   This transition is possible as the topological modes stay at fixed energy as the gap is tuned by changing the parameters of $A(\omega=0)$.  In terms of the energy bands of the wire this transition is  understood as follows:  Since $A(\omega=0)$ has fixed eigenvalues of magnitude $2$, $\omega=0$ corresponds to a band edge of the energy bands of the isolated wire. The tuning of the gap between  $\lambda_{k^*}$ and the other eigenvalues, corresponds to shifting the position of the other energy bands of the wire with respect to the band edge at $\omega=0$. Then, if $\omega=0$ does not lie within any other energy band of the wire, the behavior of the conductance is subdiffusive. If $\omega=0$ lies within the other energy bands of the wire, then their ballistic contribution dominates the behavior of the conductance.   
  
  The topology of $A(\omega)$ is not necessary for the transition and can in principle occur with a topologically trivial $A(\omega)$ (See appendix \ref{app:toy}). However, simple 2D models of such systems  may be difficult to construct.
  We refer to the subdiffusive behavior that occurs with topologically trivial and  non-trivial $A(\omega)$ as topological subdiffusive behavior and trivial subdiffusive behavior, even though the underlying cause for the two is the same. The reason for such a distinction is that the topological subdiffusive behavior is protected from weak disorders in $A(\omega)$. However, a fine tuning of the Fermi level around $\omega=0$ is required to see the subdiffusive behavior as the disorder may change the energy of the topological modes by some small amounts. Basically, the subdiffusive behavior enjoys the same topological protection as the modes themselves.
  
    In the  topologically trivial parameter regimes of $A(\omega)$, since the topological modes are absent, a usual transition from ballistic to exponential decay in the behavior of the conductance takes place as the gap strength is tuned across the value $4$. This is less interesting as this transition occurs because $\omega=0$ lies within the energy bands of the isolated wire and out of the energy bands of the isolated wire for the ballistic and exponential decaying conductance, respectively. 
  
	The transition points of the two transitions are equally interesting even if the exponential  transition is of less interest than the subdiffusive transition.   The reason is that at these points the bands of the bulk spectrum of $A(\omega)$ touch the value $2$ and therefore become trivially subdiffusive eigenvalues.  Their contribution to the conductance should therefore be taken into account as their interplay with the topological subdiffusive eigenvalues/exponential eigenvalues, cross terms in Eq.~(\ref{eq:condkk}), could lead to different behaviors of the conductance. This is illustrated in the next section where we numerically compute conductance of a simple model and  find that as $N_y\rightarrow \infty$, the conductance shows a $1/N_x^3$ subdiffusive behavior at the transition points. We also verify the existence of the topological subdiffusive  behavior for this model.

\section{Ballistic to subdiffusive transition for a two-dimensional Model of SSH-chains}
\label{sec:sshmdl}
 We  choose $H_1=\eta I$, and $H_0=2\eta I+H_{SSH}$, where $H_{SSH}$ is the Hamiltonian matrix of the SSH chain given by,
\begin{equation}
	H_{SSH}[y,y']= t_1\delta_{y,y'-1}+t_2\delta_{y,y'+1},
\end{equation}
 where $t_1,t_2>0$ are the hopping parameters of the chain. So, we are basically connecting SSH chains to each other via a simple hopping matrix to form the 2D wire. The spectrum of the isolated wire at finite $N_y$, but $N_x\rightarrow\infty$ is given by diagonalizing the matrix $2 \eta I+H_{SSH}+2\eta\cos(k_x)$ at each value of the wave vector $k_x\in (-\pi,\pi)$. Since, for large $N_y$, $H_{SSH}$ has two eigenvectors of zero energy in the topological regime ($t_2>t_1$), which are gapped from the bulk spectrum with the gap $|t_2-t_1|$. Correspondingly, $A(\omega=0)$ also has fixed eigenvalues at $2$, with the gap strength of $|t_2-t_1|/\eta$.  We, therefore, expect the subdiffusive transition in the conductance at $\omega=0$ for large  $N_y$ with the transition point given by $t_2=t_1+4\eta$. For all $t_2>4\eta+t_1$,  the conductance is expected to scale subdiffusively as $1/N_x^2$ while for $t_2<4\eta+t_1$ some of the eigenvalues of $A(\omega=0)$ will have magnitude less than $2$, and correspondingly the conductance will show a ballistic behavior with $N_x$. Note that the subdiffusive phase is deep within the topological phase of the SSH chain ($t_2>t_1$), and since the topological modes are necessary to see subdiffusive behavior, such a transition is absent in the parameter regimes corresponding to the topologically trivial phase of the SSH chain.  In the trivial phase $(t_1>t_2)$, we expect a transition from ballistic to  exponential  behavior of the conductance with the transition point $t_1=t_2+4\eta$.
 
  To understand the subdiffusive transition in terms of the energy bands of the isolated wire, we note that $\omega=0$ lies at the edge of the energy band given by $2\eta+2\eta\cos(k_x)$ of the isolated wire.   So, we are looking at the conductance with the Fermi level at the band edges of the isolated wire. Then a subdiffusive or ballistic behavior occurs depending on whether or not $\omega=0$ lies within other energy bands of the isolated wire. If $\omega=0$ does not lie within the other energy bands of the isolated wire, the dominant contribution is the subdiffusive  contribution from the band edge  at $\omega=0$. However, if  $\omega=0$ lies within the other bands of the isolated wire, the ballistic contribution due to the other bands dominates the behavior of the conductance. In the topologically trivial regime of the SSH chain, the topological modes are absent and therefore $\omega=0$ is no longer a band edge, so the subdiffusive behavior is replaced by an exponential behavior.
 
 \subsection{Numerical results}
 \begin{figure}
 	\centering
 	\subfigure[$~t_1=1, t_2=6>4+t_1$]{
 		\includegraphics[width=0.23\textwidth]{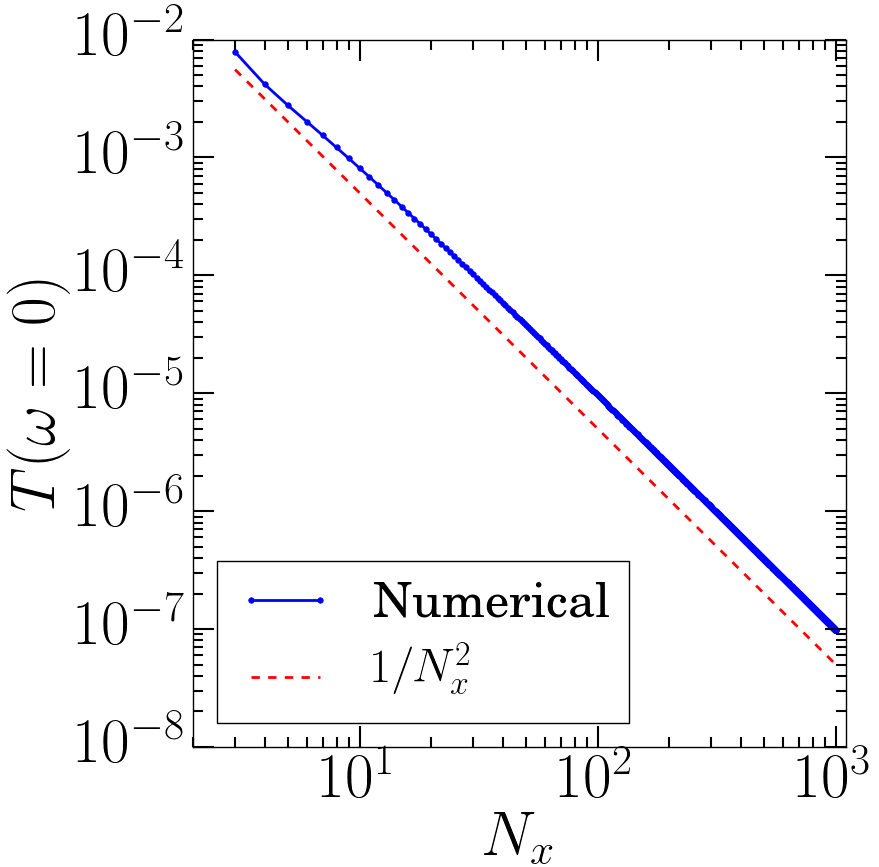}}
 	\subfigure[$~t_1=1, t_2=6>4+t_1$]{
 		\includegraphics[width=0.23\textwidth]{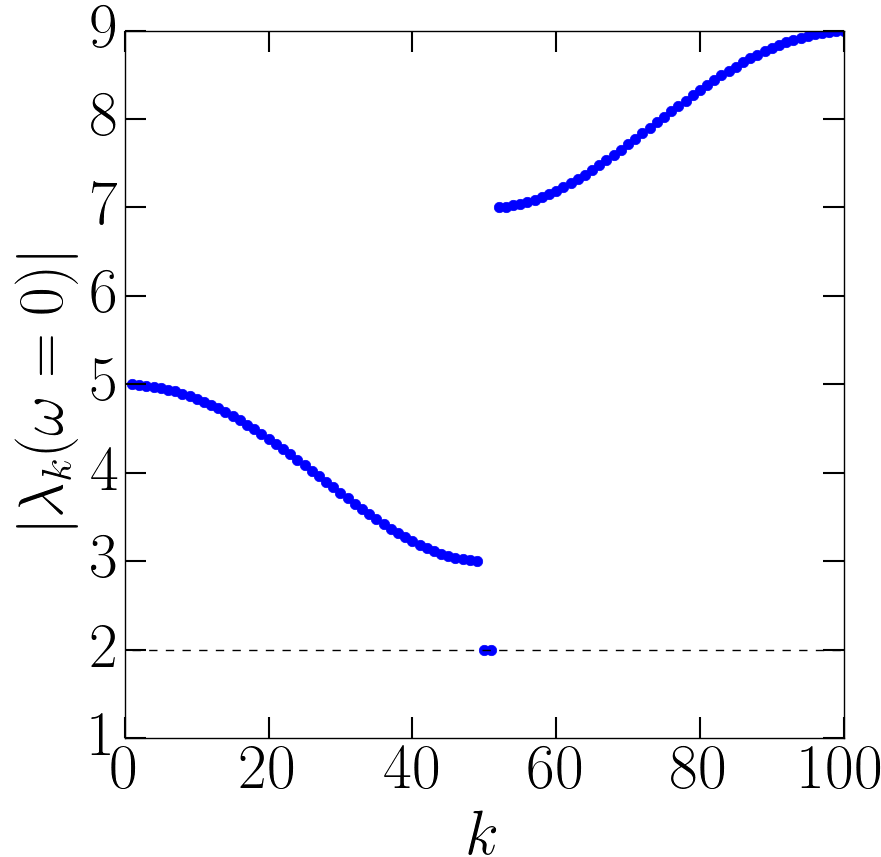}}
 	\subfigure[$~t_1=1, t_2=4<4+t_1$]{
 		\includegraphics[width=0.23\textwidth]{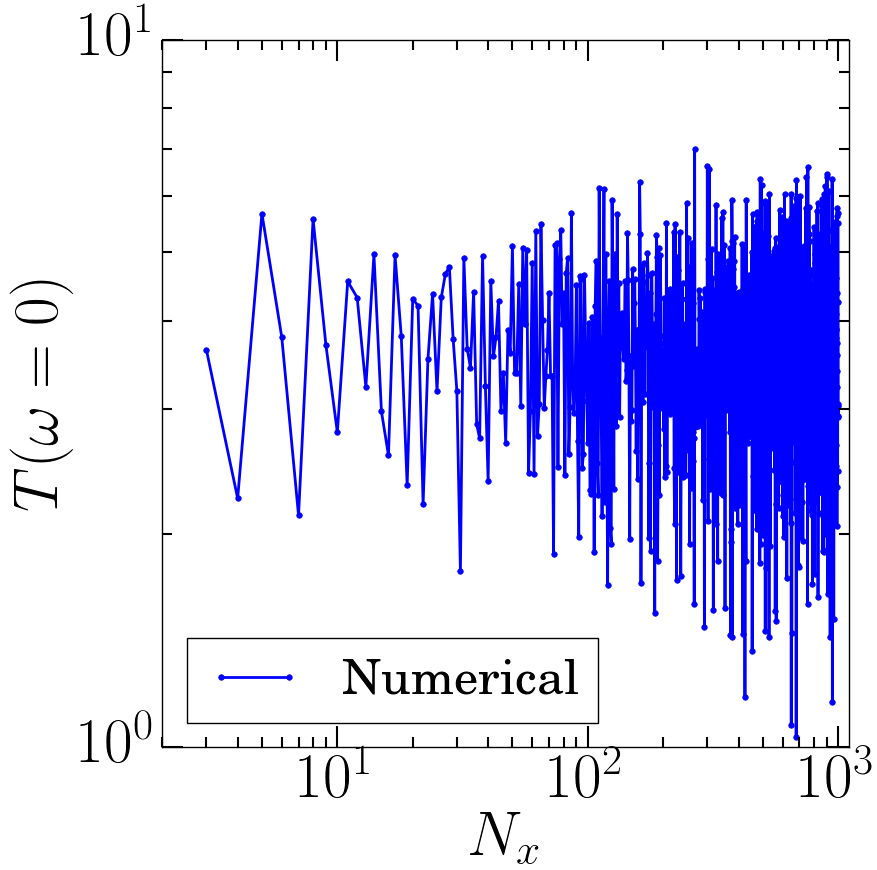}}
 	\subfigure[$~t_1=1, t_2=4<4+t_1$]{
 		\includegraphics[width=0.23\textwidth]{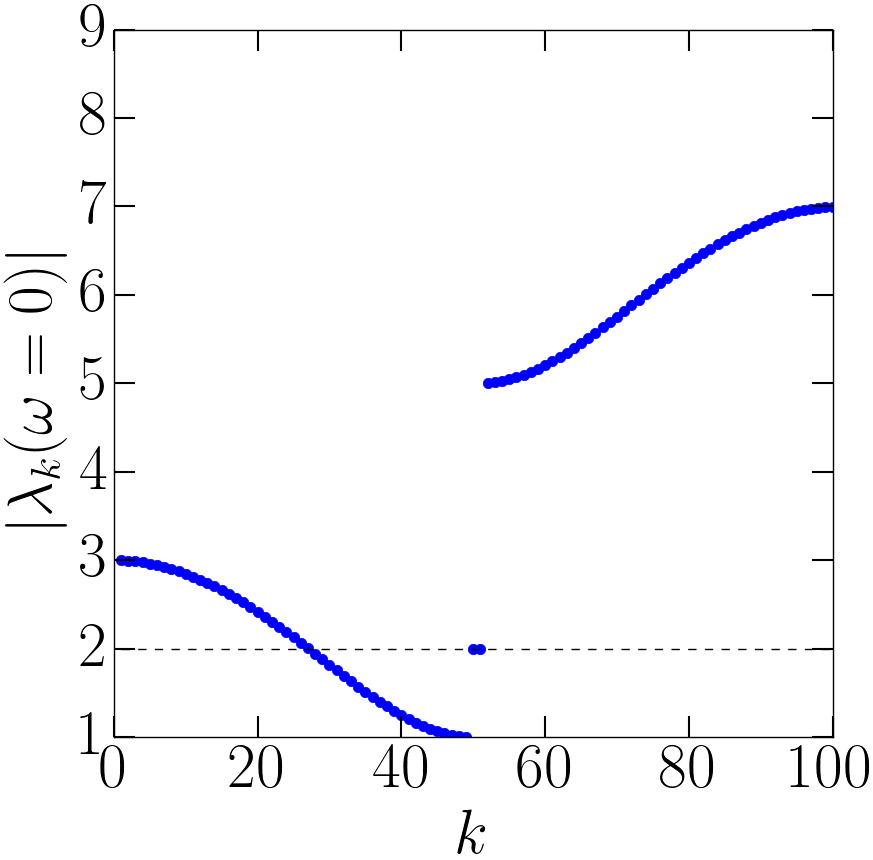}}
 	\caption{Conductance plots, \textit{log-log scale}, in the topological regime of the SSH chain with $\eta=1$, $\eta_c=0.5$, $\eta_{bx}=\eta_{by}=2$, and $N_y=100$. (a) and (c)   corresponds to the subdiffusive parameter regime, and the ballistic parameter regime, respectively. (b) and (d) are the corresponding plots of the absolute value of the spectrum of the matrix $A(\omega=0)$. We see that the numerical results show the transition from ballistic to subdiffusive behavior of the conductance as expected from the theory.}
 	\label{condpltssh}
 \end{figure}
 \begin{figure}
 	\centering
 	\subfigure[$~t_2=1, t_1=5.2>4+t_2$]{
 		\includegraphics[width=0.23\textwidth]{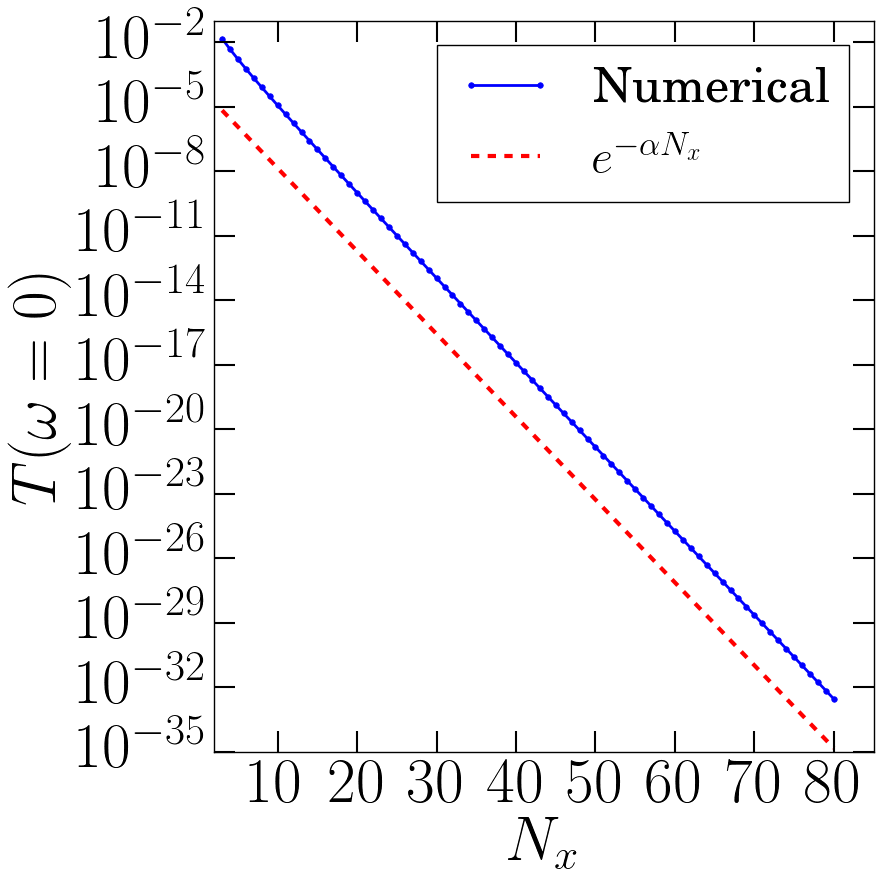}}
 	\subfigure[$~t_2=1, t_1=4<4+t_2$]{
 		\includegraphics[width=0.23\textwidth]{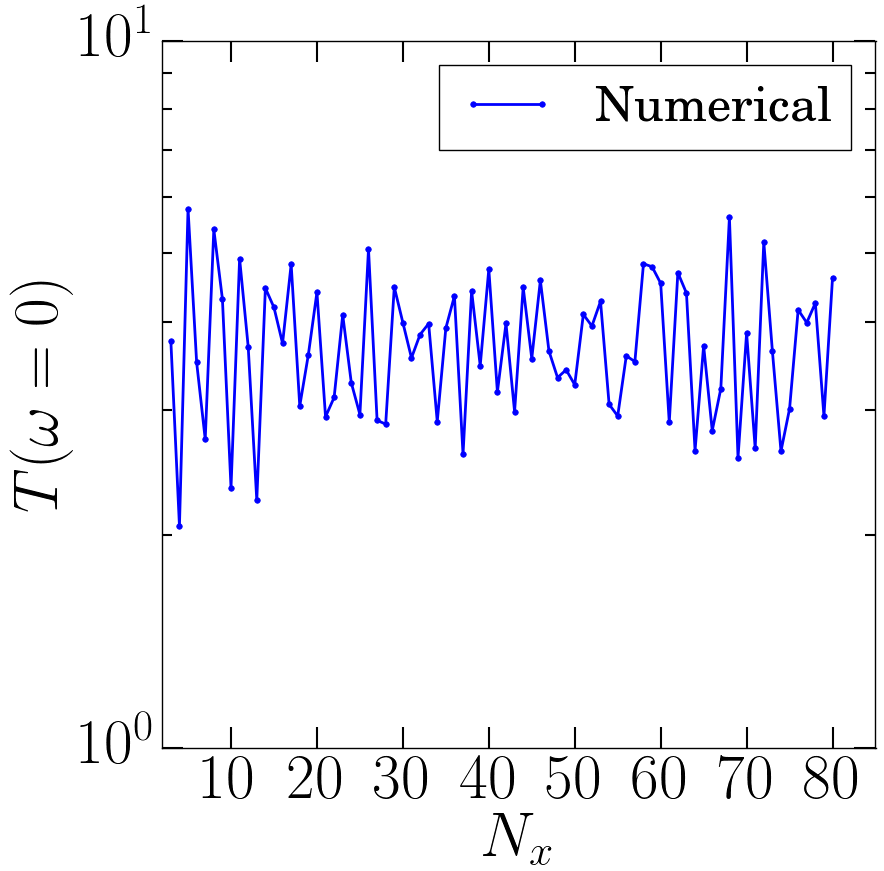}}
 	\caption{ Conductance plots, \textit{log-linear scale}, in the trivial regime of the SSH chain with $\eta=1$, $\eta_c=0.5$, $\eta_{bx}=\eta_{by}=2$, and $N_y=100$. (a) and (b)   correspond to the parameter regimes of exponential decay, and the ballistic behavior of the conductance, respectively. The numerical results show the transition from ballistic to exponential decay. $\alpha$ is defined in Eq.~(\ref{eq:exp_cond}). }
 	\label{trcondpltssh}
 \end{figure}
 
 \begin{figure}
 	\centering
 	\subfigure[$~t_1=1, t_2=6>4+t_1$]{
 		\includegraphics[width=0.4\textwidth]{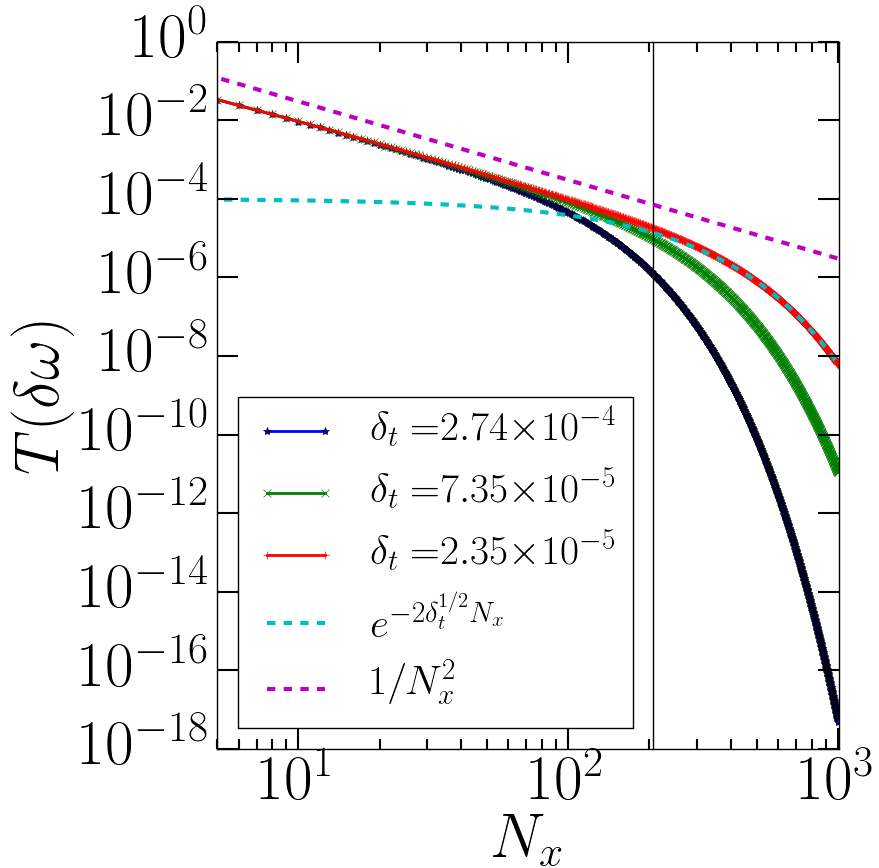}}
 	\caption{Conductance at different values $\delta_t=\delta-\delta\omega $ in presence of the onsite disorder in the chemical potential of the SSH chain. $\delta_t$ is tuned by tuning the chemical potential $\delta\omega$. The chemical potential at each site is chosen randomly from the interval $(2-\delta\mu, 2+\delta\mu)$,  $\delta\mu=0.1$. We see that as the Fermi level is finely tuned, $1/N_x^2$ behavior is observed till system sizes of order $N_x\sim 1/\delta_t^{1/2}$. This is seen from the red curve where the vertical line is drawn at $N_x=1/\delta_t^{1/2}$. The conductance shows a behavior of  $1/N_x^2$ for $N_x<<1/\delta_t^{1/2}$ and $e^{-2\delta_t^{1/2}N_x}$ for $N_x>>1/\delta_t^{1/2}$.  Parameter values-- $\eta=1$, $\eta_c=0.5$, $\eta_{bx}=\eta_{by}=2$, and $N_y=200$.  }
 	\label{dplot}
 \end{figure}
  For numerical computations we set $\eta=1$ and take the intra-chain reservoir Hamiltonian, $\tilde H_0$ to just allow nearest neighbor hoppings with hopping strength, $\eta_{by}$. Using this choice of $\tilde{H}_0$, a numerical computation of the conductance at  $\omega=0$ in the topological and trivial parameter regimes of the SSH chain is shown in Fig.~\ref{condpltssh} and Fig.~\ref{trcondpltssh}, respectively. We also show the corresponding plots of the absolute value of the  spectrum of $A(\omega=0)$ in Fig.~(\ref{condpltssh}) which we do not show for Fig.~(\ref{trcondpltssh}) as they look same as in Fig.~(\ref{condpltssh}) except that the eigenvalues corresponding to the topological modes, lying on the dotted line, are absent. From Fig.~(\ref{condpltssh}a), we see that the conductance goes down as $1/N_x^2$ for $t_2>4\eta+t_1$ as expected from the theory. The corresponding plots for the spectrum in Fig.~(\ref{condpltssh}b) has all eigenvalues of magnitude greater than $2$ except for the topological modes which lie on the dotted line at the value $2$. In Fig.~(\ref{condpltssh}c), a ballistic behavior of the conductance is observed as $t_2<4\eta+t_1$. The ballistic contribution comes from   the eigenvalues in Fig.~(\ref{condpltssh}d) which are below the dotted line. In the trivial parameter regime of the SSH chain,  we see an exponentially decaying  behavior (Fig.~(\ref{trcondpltssh}a)) and ballistic behavior (Fig.~(\ref{trcondpltssh}b)  of the conductance for $t_1>4\eta+t_2$ and $t_1<4\eta+t_2$, respectively. The subdiffusive behavior is replaced by the exponentially decaying  behavior as the topological modes are absent in this parameter regime.  

If a small disorder is added to the SSH chain, we expect the subdiffusive behavior to persist. However, the small amounts of change in the energy of the topological modes should be compensated by a small shift in the Fermi level.  To understand this, let us assume that the disorder shifts the energy of the topological modes of $A(\omega=0)$ by a small amount $\delta$, and let $k=k^*$ correspond to this eigenvalue of $A(\omega=0)$. We then have,
\begin{equation}
p_{N_x}^{k^*}=\frac{1}{\sin(q^{k^*})}\sin(q^{k^*}(N_x+1)),
\end{equation}  
where, $q^{k^*}=\arccos\frac{2+\delta}{2}$.  In order to understand the behavior at finite $N_x$, we  consider the expansion of $p_{N_x}^{k^*}$ in terms of $\delta$. We have,
\begin{equation}
q^{k^*}=\arccos\frac{2+\delta}{2}=i(\delta^{1/2}-\frac{1}{4!} \delta^{3/2}+...),
\end{equation}
and, therefore,
\begin{equation}
p_{N_x}^{k^*}=\frac{\sinh[\tilde \delta-\frac{1}{4!} \frac{\tilde\delta^{3}}{N_x^2}+...]}{\sinh[\frac{\tilde \delta}{N_x}-\frac{1}{4!} \frac{\tilde\delta^{3}}{N_x^3}+...]},\label{eq:sinser}
\end{equation}
  \begin{figure}
	\centering
	\subfigure[$~t_2=1, t_1=5=4+t_1$]{
		\includegraphics[width=0.23\textwidth]{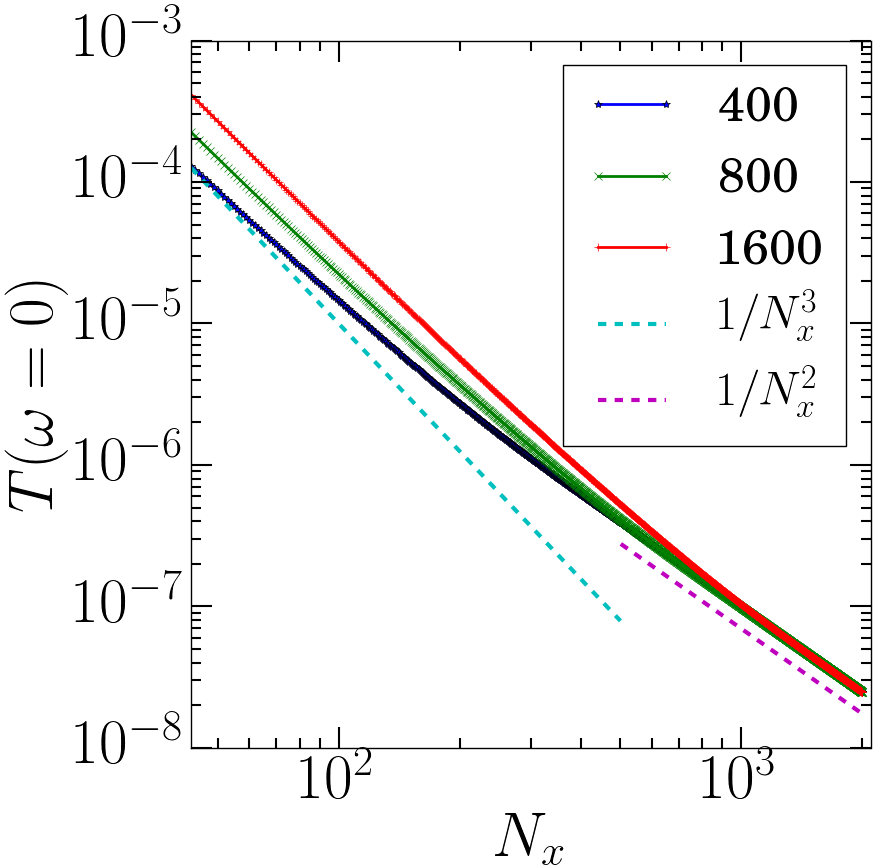}}
	\subfigure[$~t_2=1, t_1=5=4+t_2$]{
		\includegraphics[width=0.23\textwidth]{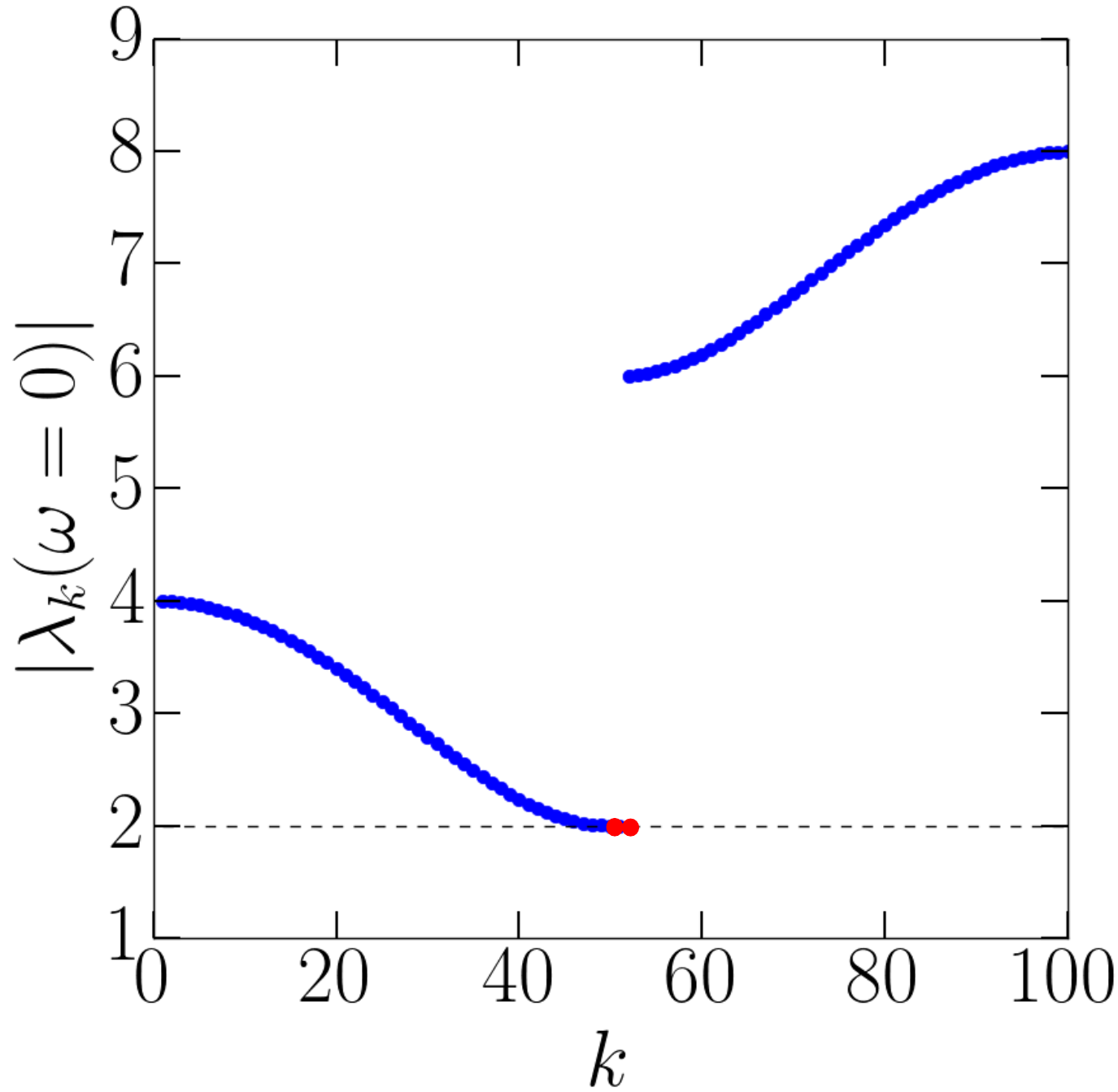}}
	\subfigure[$~t_2=1, t_1=5=4+t_1$]{
		\includegraphics[width=0.23\textwidth]{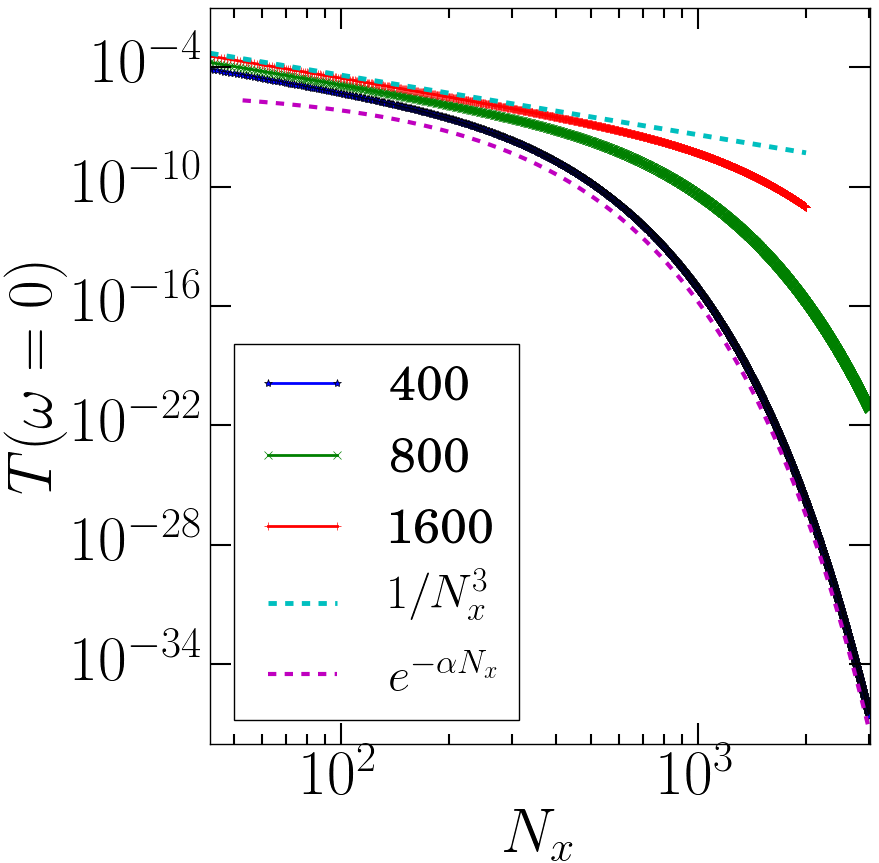}}
	\subfigure[$~t_1=1, t_2=5=4+t_1$]{
		\includegraphics[width=0.23\textwidth]{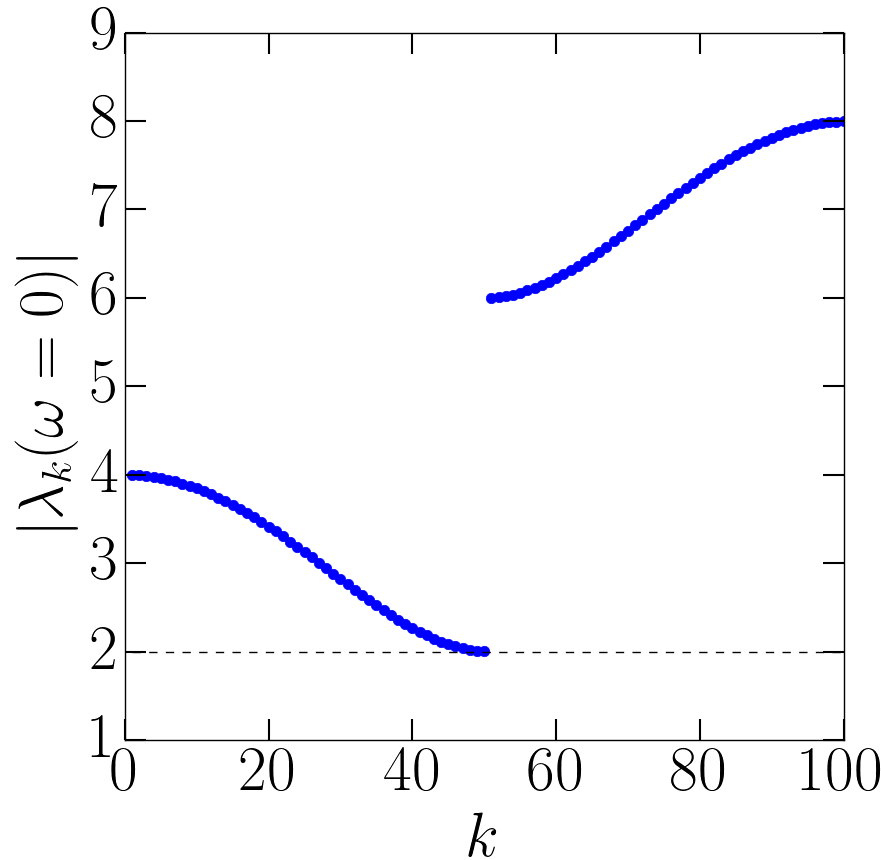}}
	\caption{Conductance, \textit{log-log scale}, and absolute value of the spectrum of $A(\omega=0)$ at the transition points of the subdiffusive (a,b) and exponential (c,d) transition, respectively. We see a scaling of $1/N_x^3$ as $N_y\rightarrow\infty$ for both  transition points, and at finite $N_y$ we observe a crossover from $1/N_x^3$ to  $1/N_x^2$ and to $e^{-\alpha N_x}$ at the subdiffusive and exponential transition points, respectively. The $1/N_x^3$ arises from the contributions of trivially subdiffusvie eigenvalues (blue dots touching $2$) apart from the topological subdiffusive eigenvalues (red dots at $2$). Other parameter values are $\eta_c=0.5$ and $\eta_{bx}=\eta_{by}=2$ and $\alpha$ is defined in Eq.~(\ref{eq:exp_cond}).}
	\label{tpcondplts}
\end{figure}

where $\tilde{\delta}=\delta^{1/2}N_x$, and we replaced $N_x+1$ by $N_x$. We see that if $\tilde \delta<<1$, then the leading contribution in Eq.~(\ref{eq:sinser}) is given by $p_{N_x}^{k^*}\sim N_x$. This implies that the conductance $T(\omega=0)\sim 1/N_x^2$. We therefore see that at a finite $N_x$, the conductance shows a behavior of $1/N_x^2$  provided $\delta<<1/N_x^2$. On the other hand, if $\tilde \delta>>1$, then the leading contribution in Eq.~(\ref{eq:sinser}) is given by $p_{N_x}^{k^*}\sim e^{\delta^{1/2}N_x}$. This leads to exponentially decaying or ballistic behavior of the conductance as $e^{-2\delta^{1/2}N_x}$ for $\delta>0$ or $\delta<0$, respectively. If the Fermi level $\omega$ is shifted to $\delta\omega$ from $\omega=0$, then the total shift in the eigenvalue corresponding to the topological mode of $A(\omega)$ is given by, $\delta_t=\delta-\delta\omega$. We show in Fig.~(\ref{dplot}) the behavior of the conductance at different values of $\delta_t$ for the SSH chain with an onsite chemical potential disorder. The vertical line in the plot is drawn at $N_x=1/\delta_t^{1/2}$, where $\delta_t$ corresponds to the red curve of the plot. The numerical calculations agree with $1/N_x^2$ and $e^{-2\delta_t^{1/2}N_x}$ for $\delta_t<<1/N^2_x$ and $\delta_t>>1/N^2_x$, respectively. 

 We now present the results at the transition points of the two transitions. The asymptotic behaviors of conductance with the wire size are shown in Fig.~\ref{tpcondplts}. We   observe a crossover from $1/N_x^3$ to  $1/N_x^2$ at the subdiffusive transition point and to $e^{-\alpha N_x}$  at the exponential transition point, respectively. The crossover region shifts towards larger values of $N_x$ as $N_y$ is increased. Therefore in the limit of $N_y\rightarrow\infty$, the numerical  results indicate a subdiffusive scaling of $1/N_x^3$ for both the transition points. A careful study of the conductance at the transition points is required to explain these behaviors.  We expect that these interesting behaviors arise due the interplay, cross terms in Eq.~(\ref{eq:condkk}), of the contribution of the three types of eigenvlaues of $A(\omega=0)$, namely topologically subdiffusive, trivially subdiffusive, and exponential eigenvalues. The trivially subdiffusive eigenvalues  exist at the transition points as at these point the edges the bulk bands of $A(\omega)$ also touch the value $2$. We  see this from plots of the spectrum at the transition points in Fig.~(\ref{tpcondplts}b) and Fig.~(\ref{tpcondplts}d) for the subdiffusive and exponential transition point, respectively.
 

\section{Conclusion}
\label{sec:concl}
In conclusion, we investigated the possibility of universal subdiffusive transport in a 2D fermionic wire defined on a square lattice. We find that such a behavior  occurs  at the values of the Fermi level, $\omega$,  where the matrix $A(\omega)=(-\omega+H_0)H_1^{-1}$ has at least one eigenvalue of magnitude $2$ while all other eigenvalues are of magnitude greater than $2$. We have shown that this implies that $\omega$ lies at the edge of an energy band of the isolated wire that does not lie within any other energy band.

We pointed out that the condition for observing the subdiffusive behavior can be simply satisfied by choosing $A(\omega)$ such that it has a non-trivial bulk topology. For such a case, the conductance shows a ballistic to subdiffusive transition asymptotically with respect to $N_x$.  The transition occurs as the positions of the other bands of the isolated wire are tuned with respect to the band edge where the Fermi level is set. Since the subdiffusive behavior arises due to the topological modes, it enjoys the same topological protection as the modes themselves. The transition point of the ballistic to subdiffusive behavior lies inside  the parameter regimes where $A(\omega)$ has a nontrivial topology. In parameter regimes corresponding to the topologically trivial phase of $A(\omega)$, the conductance shows a usual transition from ballistic to exponentially decaying behavior with respect to $N_x$.  

Interestingly, at the transition points corresponding to subdiffusive and exponential transitions  different behaviors of the conductance with respect to $N_x$ may arise for finite and thermodynamically large $N_y$. This is because at these points the edges of the bulk bands of $A(\omega)$ become subdiffusive and their interplay with the topologically subdiffusive modes/exponential modes may lead to different  $N_x$ dependence of the conductance at finite and thermodynamically large $N_y$.

We illustrate the theoretical predictions in a simple model by choosing $H_1=\eta I$ and $H_0$ to correspond to the hopping matrix of the SSH chain with a constant chemical potential. The chemical potential is added so that the eigenvalues corresponding to the topological modes of $A(\omega=0)$ lie at the value $2$. For this model,  the transition points of the subdiffusive and exponential transitions are given by $t_1=t_2+4\eta$ and $t_2=t_1+4\eta$, respectively. $t_1$, and $t_2$ are the hopping parameters of the SSH chain. We  computed the conductance of this model  numerically and  the  results  indicate the following for the conductance at $\omega=0$ for this simple model:
\begin{itemize}
	\item A transtion from ballistic to subdiffusive behavior ($1/N_x^2$) at any finite but large $N_y$ inside the topological regime. We  looked at effects of disorder on the subdiffusive behavior and showed that it persists till $N_x$ of the order of $1/\delta^{1/2}$, where $\delta$ is the shift in the energy of the topological modes of $A(\omega)$ due to the disorder.
	\item  A transition from ballistic to exponentially decaying behavior at any finite  $N_y$ inside the topologically trivial regime. 
	\item For finite $N_y$, the conductance shows a crossover from $1/N_x^3$ to $1/N_x^2$ at the subdiffusive transition point and to $e^{-\alpha N_x}$ at the exponential transition point, respectively. However,  as $N_y$ is increased the crossover region happens for larger values of $N_x$. Therefore, the subdiffusive behavior  of $1/N_x^3$  persists in the limit $N_y\rightarrow\infty$.
\end{itemize}
The results from the numerical computations of the conductance agree well with the theory away from the transition points, but a careful study of the conductance at the transition points is required to explain its behavior at these points. 

In this paper, we only looked at a specific example of  a Hermitian $A(\omega)$ having nontrivial topology. However, it may be possible to engineer models with non-Hermitian $A(\omega)$ where similar transitions in the behavior would occur, as pseudo Hermitian  matrices have been known to have non-trivial topology~\cite{lee2016anomalous,xu2017weyl,hu2011absence,schomerus2013topologically,lieu2018topological}.

\section{Acknowledgements}
I thank Marko \v{Z}nidari\v{c}, Kevin Kavanagh, and Ja\v{s} Bensa for useful discussions. The computations for the largest system size were done on the TETRIS cluster of ICTS. I also acknowledge support by Grant No. J1-4385 from the Slovenian Research Agency.

\bibliography{biblio}

\appendix

\section{General Conditions for Ballistic, Subdiffusive and Exponential Behaviour}

\label{app:gencodn}
Let $\lambda_k=a_k+i b_k$, where $a_k$ and $b_k$ are real, be the eigenvalues of the matrix $A(\omega)$. Then,
\begin{align}
q^k&=\arccos(\frac{a_k+ i b_k}{2})\\&=\cosh^{-1}{\xi_-(a_k, b_k)}+i \cosh^{-1}{\xi_+(a_k+b_k)},
\end{align}
where,
\begin{align}
\xi_{\pm}(a_k,b_k)=\frac{1}{2}\left[\sqrt{\left(\frac{a_k}{2}+1\right)^2+\frac{b_k^2}{4}}\pm \sqrt{\left(\frac{a_k}{2}-1\right)^2+\frac{b_k^2}{4}}\right].
\end{align}
Let us now discuss the three cases for the behavior of the conductance   separately,
\begin{itemize}
	\item Ballistic: For the conductance to behave balistically, we must have at least one eigenvalue $\lambda_k$ for which the imaginary part of $q^k$ vanishes. Therefore, at least one of the eigenvalues should have real and imaginary parts which lie on the curve, $\xi_+(a,b)=1$, in the $ab$ plane. This equation simply implies that $b=0$ and $-2<a<2$. So, the condition for the ballistic behavior remains the same as in the case of real spectrum of $A(\omega)$.
	
	\item Exponential: For the exponential behavior, the imaginary part of all $q^k$ must survive. So, no eigen value of $A(\omega)$ should lie on the real line between $-2$ and $2$. Hence, the contribution due to the imaginary eigenvalues is always exponentially small in $N_x$. 
	\item Subdiffusive: This behavior comes from the non-diagonalizibility of the matrix, $M=\begin{pmatrix}\lambda_k && 1\\-1 &&0\end{pmatrix}$. This only happens for $\lambda_k=2$. Then the condition for the sub diffusive behavior is that no eigenvalue should lie on the real line between $-2$ and $2$ and simultaneously at least one of the eigenvalues should be of magnitude  $2$. 
\end{itemize}
It is straight forward to see if $\lambda_k(\omega)'$s are real the conditions discussed in the main text for real spectrum of $A(\omega)$ are reproduced. Note that the conditions for observing the three behaviors of the conductance are not very different from the case of real spectrum of $A(\omega)$, therefore all the complex eigenvalues of $A(\omega)$ are irrelevant.  

\section{Eigenvalues of absolute value $2$ only at the band edges of the isolated wire}
\label{app:sdproof}
To show that the edges of the energy bands of the isolated wire always contribute subdiffusively, we need to show that $A(\omega)$ has eigenvalues of absolute value $2$ at $\omega$ equal to  the band edges. We first find a correspondence between eigenvalues of the matrix  $A(\omega)$ and the energy eigenvalues of the isolated wire. The latter is given by diagonalizing the matrix, $H_0+2H_1\cos k_x$, $k_x\in (-\pi,\pi)$.  The characteristic polynomial of this matrix is given by,
\begin{align}
	&\det[H_0+2H_1\cos k_x -\epsilon]\notag\\&=\det[(A(\omega)+2\cos k_x)H_1+\omega -\epsilon],
\end{align}
where $\det[.]$ denotes the determinant of the matrix. Clearly, for the Fermi level, $\omega$, exactly equal to one of  the energy eigenvalues of the isolated wire, $\epsilon(k_x)$, the above determinant reads, $\det[A(\omega=\epsilon(k_x))+2\cos k_x]\det[H_1]$, which vanishes for  $\lambda_k=-2 \cos k_x$. Therefore, a particular energy eigenvalue of the isolated system, $\epsilon(k_x)$,  would correspond to that eigenvalue $\lambda_k$ of $A(\omega=\epsilon(k_x))$ for which $\lambda_k=-2\cos(k_x)$. We see that  the eigenvalues of $A(\omega)$ which correspond to the energies within the  bands of the system have an absolute value less than $2$ for arbitrary $H_0$ and $H_1$.  Using this correspondence,  we also see that  $\lambda_k=\pm 2$,  corresponds to $k_x=-\pi,0,\pi$. These values of $k_x$  correspond to  the band edges of the isolated wire. This is because  at these points
\begin{equation}
	\frac{\partial\epsilon(k_x)}{\partial k_x}\sim\bra{\psi_{\epsilon(k_x)}}H_1\ket{\psi_{\epsilon(k_x)}}\sin(k_x),
\end{equation}
 where $\ket{\psi_{\epsilon(k_y)}}$ is the eigenstate corresponding to $\epsilon(k_y)$, vanishes.

\section{A toy example with $[H_0,H_1]\neq 0$}
\label{app:toy}
  \begin{figure}
	\centering
	\subfigure[$~\theta=\sqrt{2}+0.1$]{
		\includegraphics[width=0.23\textwidth]{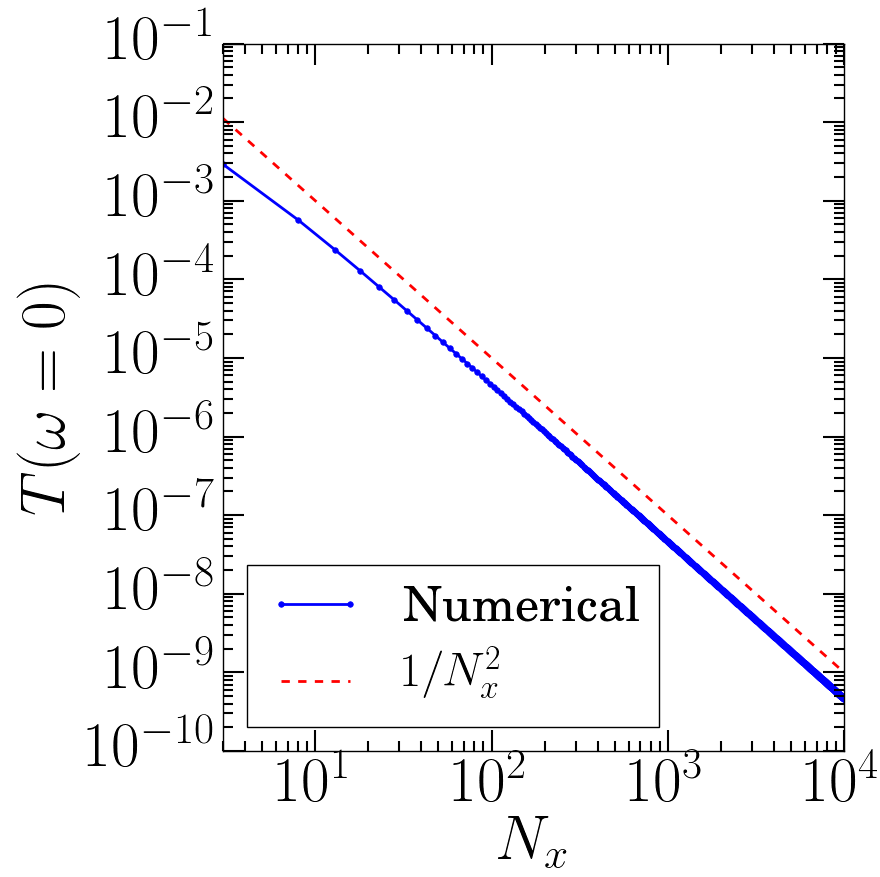}}
	\subfigure[$~\theta=\sqrt{2}+0.1$]{
		\includegraphics[width=0.23\textwidth]{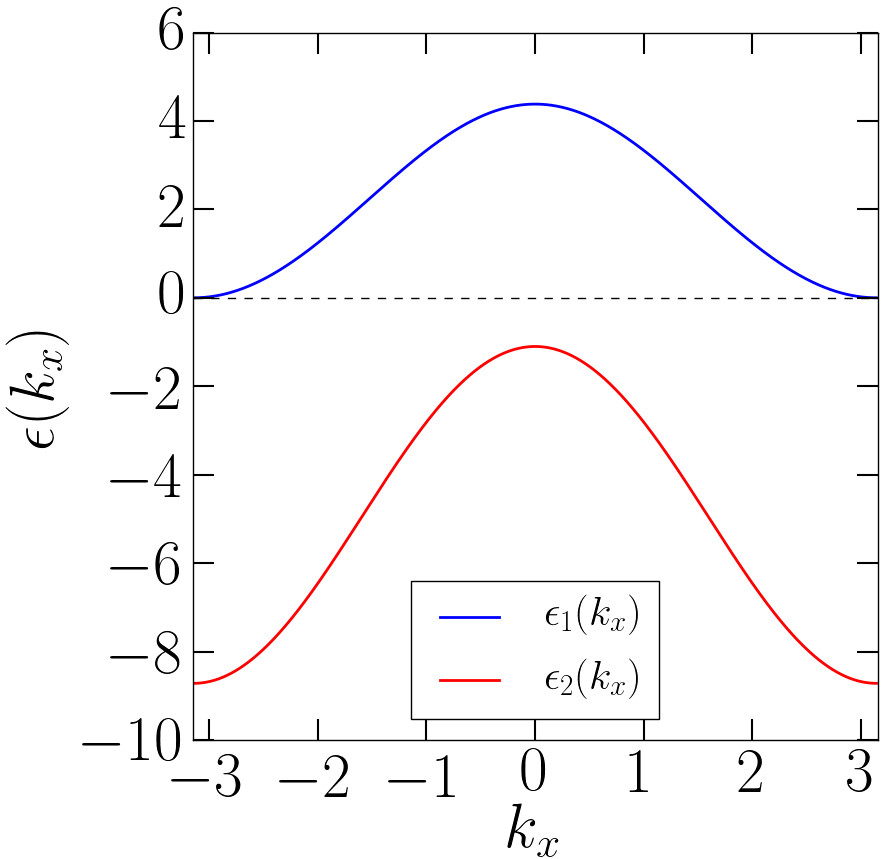}}
	\subfigure[$~\theta=\sqrt{2}-0.1$]{
		\includegraphics[width=0.23\textwidth]{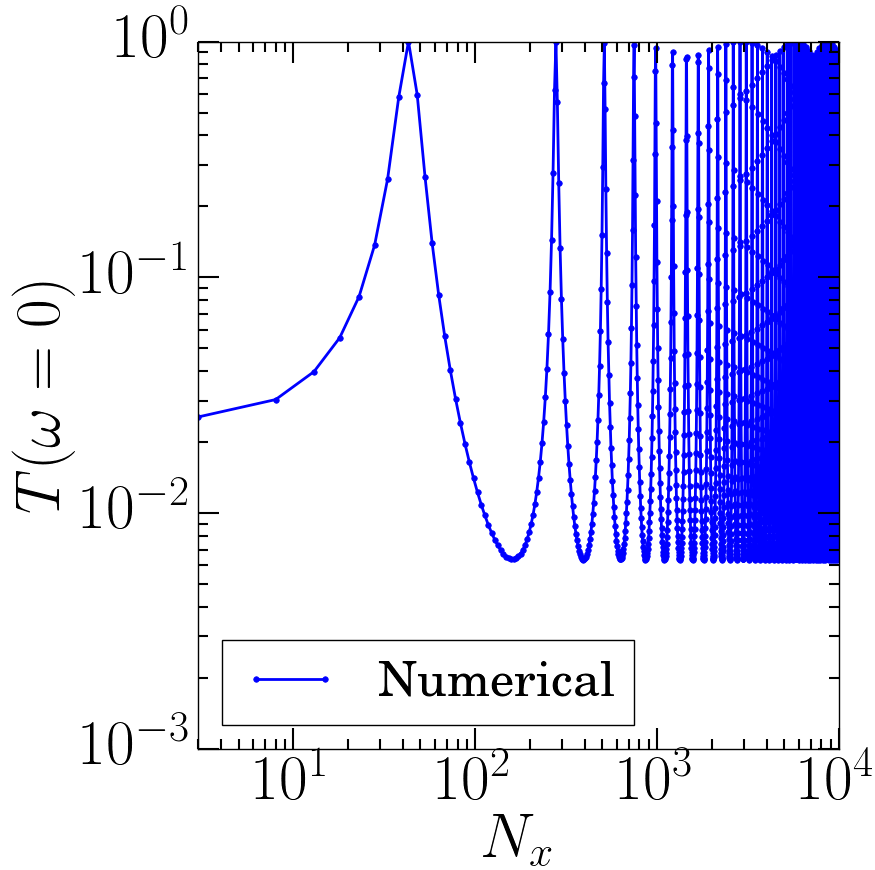}}
	\subfigure[$~\theta=\sqrt{2}-0.1$]{
		\includegraphics[width=0.23\textwidth]{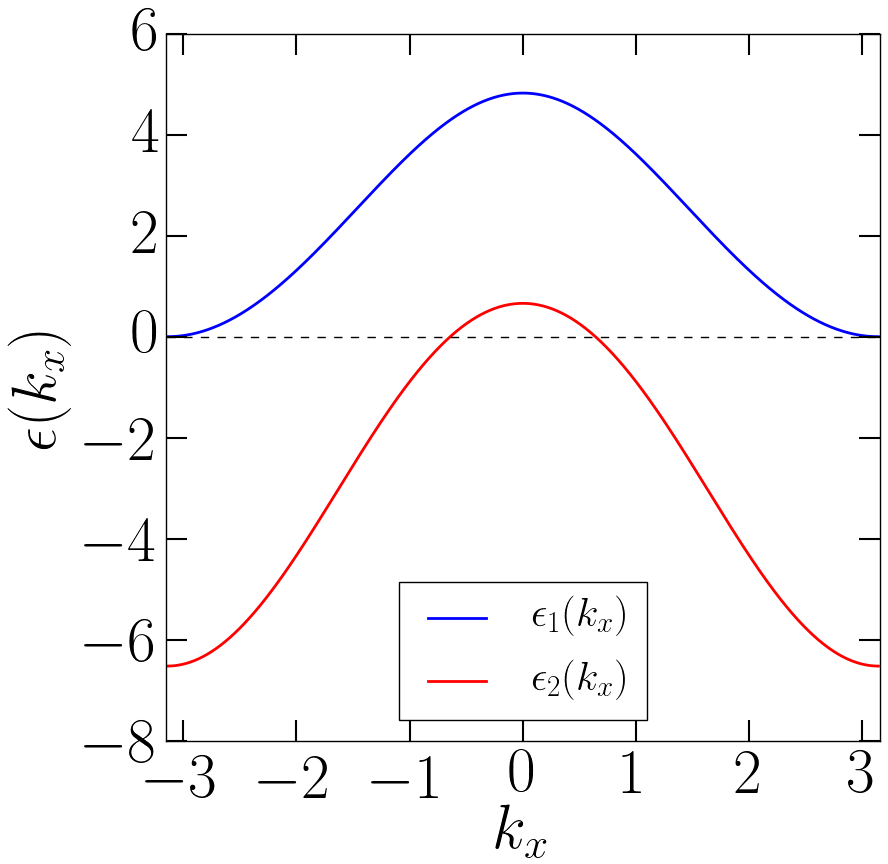}}
	\caption{ Conductance, \textit{log-log scale}, at zero Fermi level for the toy model at $\theta=\sqrt{2}\pm0.1$ and the corresponding plots for the spectrum of the wire. A transition from ballistic to subdiffusive behavior is seen as $\theta$ is tuned across the region where its absolute value is $\sqrt{2}$. Other parameter values are $\eta_c=0.5$ and $\eta_{bx}=\eta_{by}=2$. }
	\label{toyplts}
\end{figure}

To construct an example, let us work in the diagonal basis of $H_1$, with $N_y=2$. We are free to choose eigenvalues of $H_1$, we take them to be $\mu_1$ and $\mu_2$. We also fix $\omega=0$ for simplicity, and denote $A(\omega=0)$ as $A$. We note that  the pseudo Hermiticity of $A$, $A^\dagger=H_1^{-1}AH_1$, fixes it to be of the form,
\begin{equation}
A= \begin{pmatrix}
A_{11} && A_{12}\\
\frac{\mu_2}{\mu_1} A_{12}^* && A_{22}
\end{pmatrix},
\end{equation}  
where $A_{11}$ and $A_{22}$ are real. Consequently, $H_0$ is given by,
\begin{equation}
H_0=\begin{pmatrix}
A_{11}\mu_1 && A_{12}\mu_2\\
A_{12}^*\mu_2 && A_{22}\mu_2
\end{pmatrix}.
\end{equation}
Let us now use the free parameters of $A$ to get an eigenvalue of  $2$. A simple case would be to consider $\mu_1=1$ and $\mu_2=2$ and $A$ to be of the form,
\begin{equation}
A=\begin{pmatrix}
\theta && 1\\
2   && 2\frac{\theta-1}{\theta-2}
\end{pmatrix},~~\theta\neq 2.
\end{equation}
The eigenvalues of this choice of $A$ are $2$ and $\frac{2}{\theta-2}+\theta$. With this $A$ and $H_1$, $H_0$ is given by,
\begin{equation}
H_0=\begin{pmatrix}
\theta && 2\\
2 && 4\frac{\theta-1}{\theta-2}
\end{pmatrix}.
\end{equation}
We note that $A$ will always have an eigenvalue of  absolute value $2$, albeit the other eigenvalue is of magnitude less than $2$ only if $|\theta|<\sqrt{2}$. This means a ballistic to subdiffusive transition in the behavior of the conductance at zero Fermi level would occur as $\theta$ is tuned outside the region where $|\theta|<\sqrt{2}$. A numerical computation of the conductance  verifies this transition, and can be seen from  Fig.~(\ref{toyplts}a) and Fig.~(\ref{toyplts}c). Fig.~(\ref{toyplts}a)  is  for $|\theta|>2$ and Fig.~(\ref{toyplts}c) is for  $|\theta|<2$, and the two figures show a subdiffusive and ballistic behavior of the conductance, respectively. 

This transition could be understood more physically in terms of the energy eigenstates of the system. Since $N_y=2$, the wire has two energy bands given by, $\epsilon_1(k_x)$ and $\epsilon_2(k_x)$. These are given by the eigenvalues of the matrix, $H_0+2H_1\cos(k_x)$. Now as at $\omega=0$, $A(\omega)$ has  has an eigenvalue of magnitude $2$ for all values of $\theta\neq 2$,  the edge  of one of the energy bands, say $\epsilon_1(k_x)$, is  fixed at $\omega=0$. If $|\theta|<\sqrt{2}$ then  $\omega=0$ lies inside the other band, $\epsilon_2(k_x)$, and therefore its ballistic contribution dominates the transport. Outside this  range of values for $\theta$,  $\omega=0$ lies outside the other band, and therefore the only contribution is the subdiffusive  contribution of the band edge of $\epsilon_1(k_x)$. These features can be seen from the spectrum of the wire for $|\theta|>2$ and $|\theta|<2$ in Fig.~(\ref{toyplts}b) and Fig.~(\ref{toyplts}d), respectively.

\end{document}